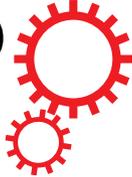

# Fast non-Abelian geometric gates via transitionless quantum driving

J. Zhang[1,2], Thi Ha Kyaw[2], D. M. Tong[1], Erik Sjöqvist[3,4] & Leong-Chuan Kwek[2,5,6,7]

A practical quantum computer must be capable of performing high fidelity quantum gates on a set of quantum bits (qubits). In the presence of noise, the realization of such gates poses daunting challenges. Geometric phases, which possess intrinsic noise-tolerant features, hold the promise for performing robust quantum computation. In particular, quantum holonomies, i.e., non-Abelian geometric phases, naturally lead to universal quantum computation due to their non-commutativity. Although quantum gates based on adiabatic holonomies have already been proposed, the slow evolution eventually compromises qubit coherence and computational power. Here, we propose a general approach to speed up an implementation of adiabatic holonomic gates by using transitionless driving techniques and show how such a universal set of fast geometric quantum gates in a superconducting circuit architecture can be obtained in an all-geometric approach. Compared with standard non-adiabatic holonomic quantum computation, the holonomies obtained in our approach tends asymptotically to those of the adiabatic approach in the long run-time limit and thus might open up a new horizon for realizing a practical quantum computer.

Fast and robust quantum gates play a central role in realizing a practical quantum computer. While the robustness offers resilience to certain errors such as parameter fluctuations, the fast implementation of designated quantum gates increases computational speed, which in turn decreases environment-induced errors. A possible approach towards robust quantum computation is to implement quantum gates by means of different types of geometric phases[1–4]; an approach known as holonomic quantum computation (HQC)[5–10]. Such geometric gates depend solely on the path of a system evolution, rather than its dynamical details.

Universal quantum computation based purely on geometric means has been proposed in the adiabatic regime, resulting in a precise control of a quantum-mechanical system[7]. Despite the appealing features, the adiabatic evolution is associated with long run time, which increases the exposure to detrimental decoherence and noise. However, this drawback can be eliminated by using non-adiabatic HQC schemes based on Abelian[8,9] or non-Abelian geometric phases[10]. The latter has been developed further in refs 11–14 experimentally demonstrated in refs 15–18 and its robustness to a variety of errors has been studied in refs 19,20.

Adiabatic processes can also be carried out swiftly by employing transitionless quantum driving algorithm (TQDA) if the quantum system consists of non-degenerate subspaces[21]. This is also known as adiabatic shortcut in the literature[22–27]. A key notion of TQDA is to seek a transitionless Hamiltonian so that the system evolves exactly along the same adiabatic passage of a given target Hamiltonian, but at any desired rate. This is achieved with the aid of an additional Hamiltonian that suppresses the energy level fluctuations caused by the changes in the system parameters.

In this report, we generalize TQDA to degenerate subspaces, where non-Abelian geometric phases are acquired after a cyclic evolution. With the help of the generalized TQDA, we propose a universal set of non-adiabatic holonomic single- and two-qubit gates. Specifically, non-Abelian geometric phases or quantum holonomies are acquired by a degenerate subspace after a cyclic evolution. TQDA-based geometric gates are realized via non-adiabatic evolution, dictated by an additional transition-suppressing Hamiltonian. We further simplify the transitionless Hamiltonian by selectively choosing geodesic path segments forming a loop in the system parameter space. This effectively reduces the complexity of the control parameters, which in turn simplifies the physical implementation of the TQDA-gates. Finally, we show how to realize the effective Hamiltonians in superconducting tunable coupling transmons.

[1]Department of Physics, Shandong University, Jinan 250100, China. [2]Centre for Quantum Technologies, National University of Singapore, 3 Science Drive 2, Singapore 117543, Singapore. [3]Department of Quantum Chemistry, Uppsala University, Box 516, Se-751 20 Uppsala, Sweden. [4]Department of Physics and Astronomy, Uppsala University, Box 516, Se-751 20 Uppsala, Sweden. [5]MajuLab, CNRS-UNS-NUS-NTU International Joint Research Unit, UMI 3654, Singapore. [6]Institute of Advanced Studies, Nanyang Technological University, 60 Nanyang View, Singapore 639673, Singapore. [7]National Institute of Education, Nanyang Technological University, 1 Nanyang Walk, Singapore 637616, Singapore. Correspondence and requests for materials should be addressed to D.M.T. (email: tdm@sdu.edu.cn) or E.S. (email: erik.sjoqvist@physics.uu.se) or L.-C.K. (email: cqtklc@nus.edu.sg)

## Results

**Degenerate TQDA.** Let us consider a quantum-mechanical system characterized by a Hamiltonian $H_0(\vec{\lambda})$ that depends on a set of control parameters $\vec{\lambda}$. For an eigenvalue $E_n(\vec{\lambda})$ of the Hamiltonian, we assume there exists a set of $m_n$ eigenstates $|\varphi_k^n(\vec{\lambda})\rangle$ ($\{k = 1, \cdots, m_n\}$) that span the proper subspace $\mathcal{M}_n$ of Hilbert space $\mathcal{H}$. When the parameter $\vec{\lambda}$ is varied from an initial value $\vec{\lambda}_i$ to a final value $\vec{\lambda}_f$, the adiabatic theorem entails that $\mathcal{M}_n$ does not mix with other subspaces in $\mathcal{H}$ when the corresponding run time tends to infinity. The states $|\widetilde{\varphi}_k^n\rangle$ satisfying the Schrödinger equation along the path in parameter space are related to $|\varphi_k^n\rangle$ as $|\widetilde{\varphi}_k^n\rangle = \sum_l |\varphi_l^n\rangle C_{lk}^n$ with $C^n$ being unitary matrices. By substituting this relation into the Schrödinger equation, we arrive at

$$C^n = \mathcal{T} \exp\left(-\int A^n dt\right),\tag{1}$$

up to a global dynamical phase. Here, $\mathcal{T}$ is a time-ordering operator, the connection $A_{kl}^n = \langle \varphi_k^n | \dot{\varphi}_l^n \rangle$ is an anti-Hermitian $m_n \times m_n$ matrix, and we have assumed the initial condition $C^n(0) = \hat{1}^n$, $\hat{1}^n$ being the $m_n \times m_n$ unit matrix. We focus on a cyclic adiabatic evolution, i.e., $\vec{\lambda}_f = \vec{\lambda}_i$. A non-trivial transformation relating the initial and final states is a quantum gate. This transformation operator, be it Abelian ($m_n = 1$) or non-Abelian ($m_n \geq 2$), is a quantum holonomy.

The transitionless Hamiltonian that exactly generates the adiabatic time evolution governed by the unitary operator $U = \sum_{n,k} |\widetilde{\varphi}_k^n\rangle \langle \varphi_k^n(0)|$, can be obtained by reverse engineering[21] yielding

$$H = H_0 + H_1 = \sum_{n,k} E_n |\varphi_k^n\rangle \langle \varphi_k^n| + i \sum_{n,k,l} (|\dot{\varphi}_k^n\rangle \langle \varphi_k^n| - A_{kl}^n |\varphi_k^n\rangle \langle \varphi_l^n|).\tag{2}$$

Since we have made no assumptions regarding the dimensionality $m_n$ of the subspaces, this result generalizes Berry's TQDA[21] to the degenerate case. From this point onwards, we denote $H_0 = \sum_{n,k} E_n |\varphi_k^n\rangle \langle \varphi_k^n|$ as the target Hamiltonian. One notes that the expectation values $\langle \varphi_k^n | H | \varphi_k^n \rangle$ and $\langle \varphi_k^n | H_0 | \varphi_k^n \rangle$ coincide, i.e., the dynamical phases acquired are the same, although the eigenstates $|\varphi_k^n\rangle$ of $H_0$ are not eigenstates of the transitionless Hamiltonian $H$. In cyclic evolution, a non-Abelian geometric phase, induced by the matrix-valued connection appearing in the additional transition-suppressing Hamiltonian, is picked up by each subspace. In the following, we demonstrate how to arrive at desired quantum gates from non-Abelian geometric phases under the degenerate TQDA.

**Fast holonomic single-qubit gates.** To implement our proposed holonomic gates, we consider a generic system consisting of four bare energy eigenstates $|0\rangle$, $|1\rangle$, $|a\rangle$, and $|e\rangle$, coupled by external oscillating fields in a tripod configuration[7]. Here, $|0\rangle$, $|1\rangle$ are qubit states; $|e\rangle$ and $|a\rangle$ are excited and auxiliary states, respectively. By assuming that all fields are resonant with the transitions, the target Hamiltonian of such a system is written as

$$H_0 = \Omega |e\rangle (f_{e0} \langle 0| + f_{e1} \langle 1| + f_{ea} \langle a|) + \text{H.c.},\tag{3}$$

where each $f_{kl}$ serves as a time-dependent control parameter for the $l \leftrightarrow k$ transition in a frame that rotates with the fields. $H_0$ has two dark and two bright eigenstates, and the former eigenstates provide a proper subspace to encode a qubit. To obtain our desired holonomic gates, we evolve the subspace along some loops in parameter space. In the standard adiabatic scheme[7], this task is accomplished by changing the control parameters slowly so that transitions between the dark and bright subspaces become negligible. In the following, we show explicitly how to speed up this process with the help of degenerate TQDA. Since the qubit information is stored in the dark subspace, there is no dynamical contribution. Therefore, the desired quantum holonomy is acquired at the end of the evolution in the same manner as in ref. 7.

We first construct a phase-shift gate $U_P = e^{i\gamma_1 |1\rangle \langle 1|}$, where the phase $\gamma_1$ depends solely on the path in the parameter space defined by the $f_{kl}$'s. This holonomic gate is implemented adiabatically with the $\Lambda$-like Hamiltonian $H_0^P = \Omega_1 |e\rangle \left(-\sin\frac{\theta}{2} e^{i\varphi} \langle 1| + \cos\frac{\theta}{2} \langle a|\right) + \text{H.c.}$, where the state $|0\rangle$ is decoupled ($f_{e0} = 0$). The Hamiltonian $H_0^P$ has a parameter-dependent dark eigenstate $|D\rangle = \cos\frac{\theta}{2} |1\rangle + \sin\frac{\theta}{2} e^{i\varphi} |a\rangle$, and two bright eigenstates. The dark state $|D\rangle$ initially coincides with $|1\rangle$ by choosing $\theta = \varphi = 0$ at $t = 0$. After completing a cyclic adiabatic evolution, this qubit state acquires the Berry phase factor $e^{i\gamma_1} = e^{-\frac{1}{2} \iint_S \sin\theta d\theta d\varphi}$, with $S$ being the surface enclosed by the path shown in Fig. 1(a). Thus, $\gamma_1$ is minus half the solid angle enclosed on the parameter sphere with polar angles $\theta$ and $\varphi$.

The shortcut to this adiabatic process is realized by adding an extra Hamiltonian term $H_1^P$ to the target Hamiltonian $H_0^P$. It is evident that $H_1^P$ relies both on the structure and rate of change of $H_0^P$. In order to simplify $H_1^P$, we propose evolving the system in three steps forming an "orange slice" on the parameter sphere along which the connection $A$ vanishes (see Fig. 1(a,b) and Supplementary information for details). First, $\varphi = 0$ while changing $\theta$ from 0 to $\pi$. During this step, terms containing $\dot{\varphi}$ and $\sin\varphi$ vanish, and $H^P = H_0^P + H_1^P$ becomes a $\Delta$-like Hamiltonian, in which the Rabi frequency between $|1\rangle$ and $|a\rangle$ is $\dot{\theta}/2$. Therefore, a $1 \leftrightarrow a$ transition is added to the system caused by the temporal change in $\theta$. In the second step, we keep $\theta = \pi$, but vary $\varphi$ from 0 to $\varphi_1$. Here, the total Hamiltonian is just a transition between $|e\rangle$ and $|1\rangle$ with detuning $\dot{\varphi}$. Thus, the detuning is caused by the temporal change in the parameter $\varphi$. Finally, we keep $\varphi = \varphi_1$, while $\theta$ is decreased to 0. During this final step, $H^P$ is again a $\Delta$-like Hamiltonian. When $\theta = 0$, $H^P$ has returned to its initial form. As shown in Fig. 1(a), the acquired geometric phase $\gamma_1$ is equal to minus the azimuthal angle $\varphi_1$, i.e., minus half the solid angle enclosed by the orange slice shaped loop. We note that both the detuning in the second step and the $1 \leftrightarrow a$ transition in the first and third steps are caused by the fast parameter change. Thus, $H_0^P$ is recovered in the adiabatic limit when the rate of the parameter change tends to zero. The steps of the control Hamiltonian needed to implement $U_P$ are shown in Table 1.

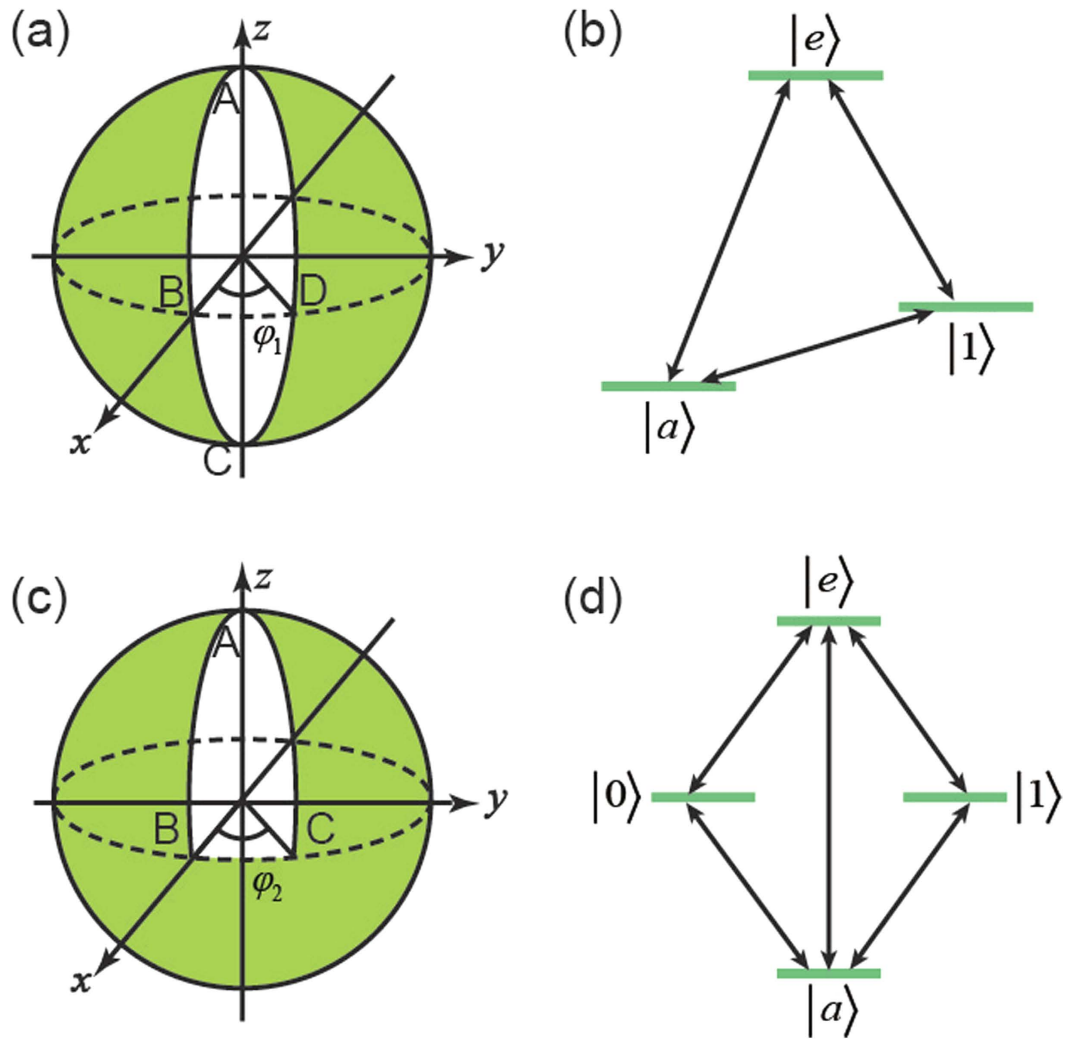

**Figure 1. Schematic diagram for the orange slice and geodesic triangle schemes.** The spheres in (**a**,**c**) represent the parameter space, where the control parameters $\theta$ and $\varphi$ are changed following a specific path. When we set $\theta = \varphi = 0$, the starting point in the parameter space is along the z axis. The parameters $\theta$ and $\varphi$ are polar angle and azimuthal angle in the parameter space, respectively. (**a**) Schematic diagram for the orange slice scheme. The state $|1\rangle$ is evolved by changing the parameters along the path A-B-C-D-A. The opening angle is $\varphi_1$. (**b**) The level structure for the first and third steps in achieving $U_P$. Transitions between all three levels are needed, which makes it a $\Delta$ model Hamiltonian. (**c**) Schematic diagram for the geodesic triangular scheme. The manifold, consisting of $|0\rangle$ and $|1\rangle$, is evolved by changing the parameters along the path A-B-C-A, while the opening angle is $\varphi_2$. (**d**) The level structure for the last step in achieving $U_B$. Transitions between all four levels are needed.

Similarly, an adiabatic gate $U_B = e^{i\gamma_2 \sigma_y}$, $\sigma_y = i(|0\rangle\langle 1| - |1\rangle\langle 0|)$ can be realized by utilizing the full tripod structure with $f_{e0} = \Omega \sin \theta \cos \varphi$, $f_{e1} = \Omega \sin \theta \sin \varphi$, and $f_{ea} = \Omega \cos \theta$. This Hamiltonian has two degenerate parameter-dependent dark eigenstates, which we take as $|D_1\rangle = \cos \theta (\cos \varphi |0\rangle + \sin \varphi |1\rangle) - \sin \theta |a\rangle$ and $|D_2\rangle = \cos \varphi |1\rangle - \sin \varphi |0\rangle$, together with two non-degenerate bright eigenstates. Similar to the adiabatic phase-shift gate, our qubit is encoded within the dark subspace. The desired holonomic gate is obtained after traversing a loop on the parameter sphere with polar angles $\theta$ and $\varphi$, resulting in $\gamma_2$ as the solid angle swept[7].

Shortcut to $U_B$ is realized by choosing a geodesic triangle path on the corresponding parameter sphere, again in order to force the connection $A$ to vanish (see Fig 1(c,d) and Supplementary information for details). The purpose of this choice is to simplify the implementation of the gate, just as the choice of an "orange slice" path simplified the implementation of $U_P$. Initially, we set $\theta = \varphi = 0$, which implies $|D_1\rangle = |0\rangle$ and $|D_2\rangle = |1\rangle$, followed by increasing $\theta$ to $\frac{\pi}{2}$, while keeping $\varphi$ constant. During this step, $H^B = H_0^B + H_1^B$ exhibits a $\Delta$-like structure involving the states $|e\rangle$, $|0\rangle$, and $|a\rangle$. Thereafter, we keep $\theta$ unchanged but rotate $\varphi$ to $\varphi_2$. $H^B$ has again a $\Delta$-like structure, but now involving $|e\rangle$, $|0\rangle$, and $|1\rangle$. Finally, $\theta$ is decreased to zero along the geodesic curve and then $\varphi$ is tuned to zero. The level structure for the last step is depicted in Fig. 1(d). Unlike the "orange slice" scheme, no detuning emerges in the "geodesic triangle" scheme. Therefore, all the photon-assisted transitions are resonant. The steps of the control Hamiltonian needed to implement $U_B$ are shown in Table. 1. We may note that all transitions between the four levels $|0\rangle$, $|1\rangle$, $|a\rangle$, and $|e\rangle$ are needed to realize an arbitrary single-qubit gate by combining $U_P$ and $U_B$.

| | Step1 | Step2 | Step3 |
|---|---|---|---|
| $U_P$ | $|e\rangle \leftrightarrow |1\rangle, |e\rangle \leftrightarrow |a\rangle, |1\rangle \leftrightarrow |a\rangle$ | $|e\rangle \leftrightarrow |1\rangle$ | $|e\rangle \leftrightarrow |1\rangle, |e\rangle \leftrightarrow |a\rangle, |1\rangle \leftrightarrow |a\rangle$ |
| $U_B$ | $|e\rangle \leftrightarrow |0\rangle, |e\rangle \leftrightarrow |a\rangle, |0\rangle \leftrightarrow |a\rangle$ | $|e\rangle \leftrightarrow |0\rangle, |e\rangle \leftrightarrow |1\rangle, |0\rangle \leftrightarrow |1\rangle$ | $|e\rangle \leftrightarrow |0\rangle, |e\rangle \leftrightarrow |1\rangle, |e\rangle \leftrightarrow |a\rangle, |0\rangle \leftrightarrow |a\rangle, |1\rangle \leftrightarrow |a\rangle$ |

**Table 1.  Transitions needed in the "orange slice" scheme and the "geodesic triangle" scheme.** The "orange slice" scheme and the "geodesic triangle" scheme are proposed to achieve the single qubit gates $U_P$ and $U_B$, respectively. Each scheme contains three steps and needed transitions for each step are listed. The symbol $|i\rangle \leftrightarrow |j\rangle$ denotes the transition between the states $|i\rangle$ and $|j\rangle$.

**Fast holonomic two-qubit gate.**     Since the combination of the gates $U_P$ and $U_B$ allows the realization of an arbitrary single-qubit gate, it remains to construct an entangling two-qubit holonomic gate to realize universal quantum computation[28,29]. A holonomic two-qubit gate can be realized in the TQDA scenario, by controlling suitable coupling parameters between two four-level tripod systems. Specifically, we note that there are four auxiliary states, $\mathcal{K} = \left\{ |e\rangle_l \otimes |e\rangle_r, |e\rangle_l \otimes |a\rangle_r, |a\rangle_l \otimes |e\rangle_r, |a\rangle_l \otimes |a\rangle_r \right\}$ from the two four-level systems (see Methods and Supplementary information for details). The key idea is to construct a $\Delta$-like structure to combine two auxiliary states from $\mathcal{K}$ and one of the two-qubit states. The target Hamiltonian of the two-qubit gate can be taken as $H_0^2 = J_1 |ea\rangle \langle 10| + J_2 |ea\rangle \langle ae| + H.c$; By using the same "orange slice" scheme as in the realization of $U_P$, a geometric phase is acquired by the state $|10\rangle$ while the other states remain unchanged. This amounts to applying an entangling gate $U_2 = e^{i\gamma_3 |10\rangle\langle 10|}$ on the two-qubit space, where $\gamma_3$ is the geometric phase obtained after the cyclic evolution.

**The single- and two-qubit gates in superconducting circuits.**     To realize our non-adiabatic schemes experimentally, we consider a superconducting tunable coupling transmon (TCT)[30] (see Fig. 2(a)), in which we assume that the transition frequency of the system and system-resonator coupling strength can be tuned independently. In the symmetric configuration, where $E_{J_a}^1 = E_{J_a}^2 = E_{J_a}^0, a = \pm$, the system Hamiltonian is described by[30,31]

$$H_0^{\text{TCT}} = \sum_{a=\pm} \left[ 4E_{C_a}(n_a - n_{g_a})^2 - E_{J_a}^0 \cos(\pi f_a) \cos(\gamma_a - 2\pi f_a) \right] + 4E_I n_+ n_-, \tag{4}$$

where $E_{C_a}$ are the charging energies of the upper and lower islands, while $E_{J_a}^i, i = 1, 2$, are the Josephson energies (see Methods for definition of the other parameters in $H_0^{\text{TCT}}$). In the limit of large $E_{J_a}^a/E_{C_a}$, the energy level splitting is observed with increase in $E_J$ as shown in Fig. 3(a), giving rise to the four energy levels $|0\rangle$, $|1\rangle$, $|a\rangle$ and $|e\rangle$ needed to implement our proposal.

When a time-dependent microwave field $\Phi(t)$ impinges onto our four-level TCT (see Figs 2(a) and 3(a) and Methods), microwave-assisted transitions occur. To investigate allowed energy level transitions, we consider a situation where there is a minimum flux entering the upper SQUID loop ($\Phi_+ \approx 0$), where a small change of flux ($\Phi_-$) is present in the lower one. In this case, the perturbation Hamiltonian with small $\Phi(t)$ becomes $H_1^{\text{TCT}}(\gamma_-, t) = \Phi^{(0)}\Gamma \cos(\omega_d t)$ with $\Phi^{(0)}$ the microwave amplitude, $\omega_d$ a carrier frequency of the incoming microwave pulse that connects two desired eigenstates $|k\rangle$ and $|l\rangle$, and $\Gamma = (E_J / \Phi_0)\cos(\pi f_-)\sin(\gamma_- - 2\pi f_-)$. In Fig. 3(b), we plot the transition matrix elements for the eigenstates of interest $|t_{kl}| = |\langle k|H_1^{\text{TCT}}|l\rangle|$ versus $E_J/E_C$ while $E_I/E_C = 0.5$ and $E_{J_+}/E_C = 100$. From this result, we conclude that a particular energy value of $E_J$ should be tuned via the external magnetic flux, threading the lower SQUID loop, in order to give rise to the desired energy level transitions. If we confine ourselves to the selected four eigenstates, the Hamiltonian of the superconducting circuit in a rotating frame can be expressed as $H^{\text{TCT}} = H_0^{\text{TCT}} + H_1^{\text{TCT}} = \sum_{k \neq l = 0,1,e,a} [\Omega_{kl} |k\rangle \langle l| + \text{H.c.}]$, where $\Omega_{kl} = \hbar \langle k|\Phi^{(0)}\Gamma|l\rangle/2$ is a complex Rabi frequency. When we invoke the "orange slice" and "geodesic triangle" schemes, $H^{\text{TCT}}$ gives rise to $H^P$ and $H^B$, respectively. We note that since our schemes relax the adiabatic constraint, the holonomic gates can be obtained at any speed. However, from a practical perspective, the desired gate time depends on details of the specific experimental setup such as the energy gaps and how fast we can vary the input microwave pulses.

To implement our two-qubit gate, we consider a superconducting architecture with two TCTs placed inside a coplanar resonator (see Fig. 2(b,c)). The direct coupling between the two transmons (labeled as $l$ and $r$) is mediated by a cavity in the resonant regime[32]. With the built-in tunability of the transmons, we realize the Hamiltonian $H^2$ as follow. We assume that the transmon levels are anharmonic and each level transition can be tuned at will by matching with the resonator frequency via ac-Stark-shift fields[33], alternatively by invoking the built-in tunability of the transmons energy levels[30] or by tuning the resonator frequency[34]. For instance, to achieve the transition between $|e_l, a_r\rangle$ and $|1_l, 0_r\rangle$, we switch on vacuum Rabi couplings $\hbar g_{1_l, e_l}(\sigma^+_{1_l, e_l} a + \sigma^-_{1_l, e_l} a^\dagger)$ to transmon $l$, and $\hbar g_{a_r, 0_r}(\sigma^+_{a_r, 0_r} a + \sigma^-_{a_r, 0_r} a^\dagger)$ to transmon $r$, where the g's are the coupling strengths between the respective transition and the single mode resonator, $\sigma^+_{j,k} = |j\rangle\langle k|$, $\sigma^-_{j,k} = |k\rangle\langle j|$, and $a^\dagger$ ($a$) is the creation (annihilation) operator of the bosonic resonator mode, respectively. All the other required transitions can also be realized in the same manner (see Supplementary information for details).

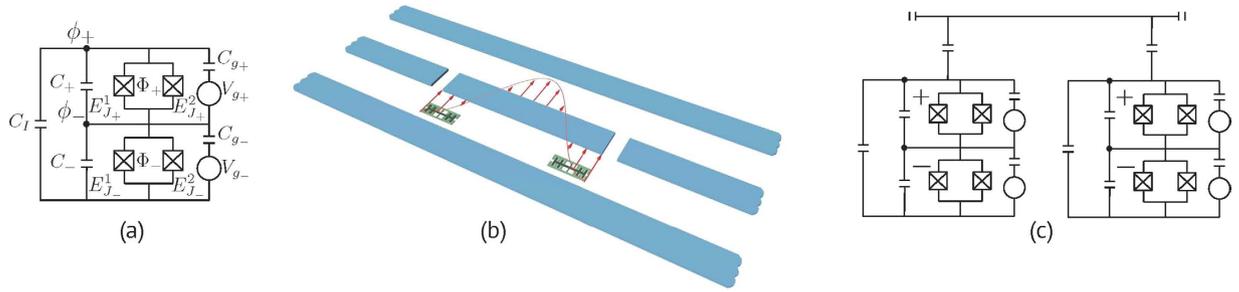

**Figure 2. Diagram for superconducting circuits. (a)** Circuit model of a tunable coupling transmon (TCT) system, which is used to realize our single-qubit gates in absence of a transmission line resonator. **(b)** A proposed setup to realize the two-qubit gate with two TCTs mediated via a resonator and **(c)** its equivalent circuit model, where the two TCTs are capacitively coupled to a coplanar resonator.

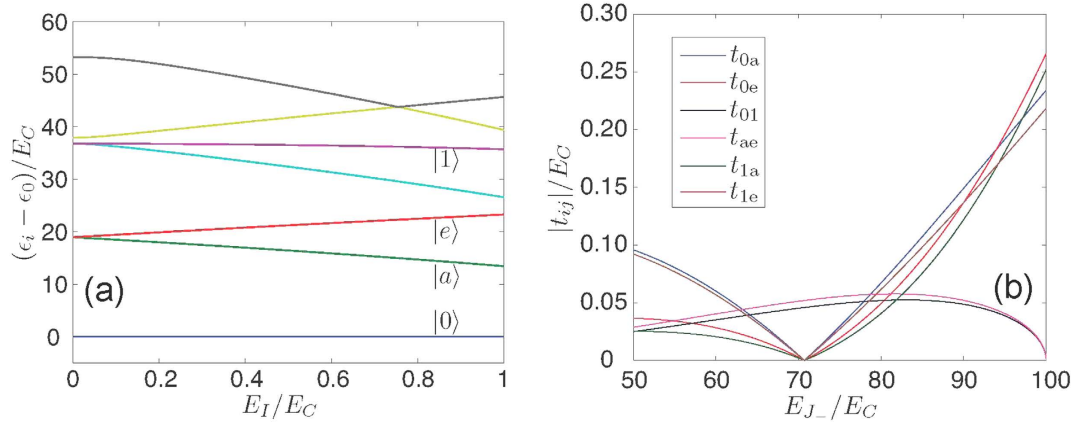

**Figure 3. Energy spectrum and transition matrix for the superconducting circuits. (a)** Energy spectrum of the TCT as a function of $E_I/E_C$ for $E_{J_\pm} = 50E_C$. We confine ourselves to the eigenstates $|0\rangle$, $|1\rangle$, $|e\rangle$ and $|a\rangle$. **(b)** Moduli $|t_{ij}|$ of the unnormalized transition matrix elements between states $|i\rangle$ and $|j\rangle$ as a function of $E_{J_-}/E_C$ for the labelled eigenstates in **(b)**, while $E_I/E_C = 0.5$ and $E_{J_+}/E_C = 100$.

## Discussion and Conclusions

We would like to put forward some remarks on the holonomic gates based on TQDA proposed here and the nonadiabatic holonomic gates proposed in ref. 10. On one hand, we notice that both schemes are genuinely non-adiabatic in that they can be performed at any desired rate. On the other hand, the holonomies obtained in the proposed TQDA approach tends asymptotically to those of the adiabatic approach in the long run-time limit[2,7]; a feature which is not shared by the scheme in ref. 10. As a consequence, the TQDA-based gates have an immediate geometric meaning in terms of solid angles enclosed in parameter space of the underlying target Hamiltonian, while no such explicit geometric meaning is present in the other scheme. This also makes our scheme explicitly robust to parameter fluctuations that preserve the solid angle. In addition, our scheme has the advantage over standard adiabatic scheme in that non-adiabatic transitions, which inevitably degrades the control in any realistic finite-time implementation of the adiabatic gates[19], are suppressed using TQDA.

In conclusion, we have proposed a non-Abelian generalization of the TQDA and show how it can be used for holonomic quantum computation. The key feature of this algorithm is to realize quantum holonomies that can be performed at arbitrary rate by applying an additional transition-suppressing Hamiltonian so that the path of the original eigensubspaces as well as the purely geometric nature of the resulting matrix-valued subspaces are preserved. Non-adiabatic TQDA-based single- and two-qubit gates are proposed in a four-level transmon and two four-level transmons coupled with a cavity, respectively. With high controllability[35,36] and scalability[37,38] of the superconducting circuit, our scheme can also be incorporated within a more general landscape where the transmons are embedded in a resonator lattice[39–43]. This scheme therefore opens up new experimental horizons towards a robust all-geometric high-speed large-scale quantum computation architecture.

## Methods

**Superconducting tunable coupling transmon.**     The four level system can be provided by a superconducting tunable coupling transmon (TCT)[30] (see Fig. 2(a)). To obtain the effective Hamiltonian $H_0^{\text{TCT}}$, we define $\gamma_+ = 2\pi(\phi_+ - \phi_-)/\Phi_0$, $\gamma_- = 2\pi\phi_-/\Phi_0$ ($\Phi_0 = h/2e$ being the superconducting magnetic flux quantum) as the phase differences on the upper and lower superconducting quantum interference device (SQUID) loops, and $n_{g_\pm} = (C_{g_\pm} V_{g_\pm})/2e$ as the dimensionless gate voltages. The tunable Josephson energies $E_{J_\pm} = E_{J_\pm}^0 \cos(\pi f_\pm)$, where $E_{J_\pm}^0 = E_{J_\pm}^1 + E_{J_\pm}^2$ and the frustration parameters $f_\pm = \Phi_\pm/\Phi_0$. $\Phi_\pm$ are external magnetic fluxes threading the upper and lower superconducting quantum interference devices (SQUIDs). In $H_0^{\text{TCT}}$, $n_\pm$ is the respective associated charge. $E_{C_\pm} = e^2/2C_\pm'$ are the charging energies of the upper and lower islands, while $E_I = -e^2/C'$ represents the interaction energy between them. By letting $C_{\Sigma_\pm} = C_I + C_\pm + C_{g_\pm}$, the effective capacitors $C_\pm'$ and $C'$ are defined by $C_\pm' = \left[C_{\Sigma_\mp}C_{\Sigma_-} - C_I^2\right]/C_{\Sigma_\mp}$ and $C' = \left[C_{\Sigma_+}C_{\Sigma_-} - C_I^2\right]/C_I$, respectively.

## Acknowledgements

J.Z. and D.M.T. acknowledge support from NSF China through Grant No. 11175105 and the National Basic Research Program of China through Grant No. 2015CB921004. T.H.K. and L.-C.K. acknowledge support from the National Research Foundation & Ministry of Education, Singapore. E.S. acknowledges financial support from the Swedish Research Council. The authors thank Guillermo Romero, Enrique Solano, Yuimaru Kubo, Patrice Bertet and Daniel Esteve for valuable discussions and helpful comments.

## Author Contributions

The authors contributed equally to this work.

## Additional Information

**Supplementary information** accompanies this paper at http://www.nature.com/srep

**Competing financial interests:** The authors declare no competing financial interests.

**How to cite this article**: Zhang, J. *et al.* Fast non-Abelian geometric gates via transitionless quantum driving. *Sci. Rep.* **5**, 18414; doi: 10.1038/srep18414 (2015).